\documentclass[12pt, a4paper]{iopart}
\usepackage{iopams}
\usepackage{makeidx}
\usepackage{theorem}
\usepackage{graphicx}
\usepackage{tabularx}
\usepackage{bbm}
\usepackage{cite}
\usepackage[justification=centering]{caption}
\newtheorem{lemma}{Lemma} 
\newtheorem{example}{Example}

\begin{document}

\title{Analogy between gambling and measurement-based work extraction}
\author{Dror~A.~Vinkler$^1$, Haim~H.~Permuter$^1$ and~Neri~Merhav$^2$}
\address{$^1$ Department of Electrical and Computer Engineering, Ben-Gurion University of the Negev, 84105, Beer-Sheva, Israel.}
\address{$^2$ Department of Electrical Engineering, Technion, Haifa 32000, Israel.}
\eads{\mailto{vinklerd@post.bgu.ac.il}, \mailto{haimp@bgu.ac.il}, \mailto{merhav@ee.technion.ac.il}}

\begin{abstract}
In information theory, one area of interest is gambling, where mutual information characterizes the maximal gain in wealth growth rate due to knowledge of side information; the betting strategy that achieves this maximum is named the Kelly strategy. In the field of physics, it was recently shown that mutual information can characterize the maximal amount of work that can be extracted from a single heat bath using measurement-based control protocols, i.e., using ``information engines". However, to the best of our knowledge, no relation between gambling and information engines has been presented before. In this paper, we briefly review the two concepts and then demonstrate an analogy between gambling, where bits are converted into wealth, and information engines, where bits representing measurements are converted into energy. From this analogy follows an extension of gambling to the continuous-valued case, which is shown to be useful for investments in currency exchange rates or in the stock market using options. Moreover, the analogy enables us to use well-known methods and results from one field to solve problems in the other. We present three such cases: maximum work extraction when the probability distributions governing the system and measurements are unknown, work extraction when some energy is lost in each cycle, e.g., due to friction, and an analysis of systems with memory. In all three cases, the analogy enables us to use known results in order to obtain new ones.
\end{abstract}


\section{Introduction}
While both work extraction from feedback controlled systems and information-theoretic analysis of gambling are old concepts, to the best of our knowledge the relation between them has not been highlighted before. This relation includes a straightforward mapping of concepts from one field to the other, e.g., measurements are analogous to side information and control protocols to betting strategies. Fundamental formulas in one field apply to the other after simple replacement of variables according to the derived mapping. This allows us to gain insights with regard to one field from known results from the other one.

The relationship between work extraction and information was first suggested by Maxwell \cite{Maxwell1871} in a thought experiment consisting of an intelligent agent, later named Maxwell's demon. The agent measures the velocity of gas molecules in a box that is divided into two parts by a barrier. Although both parts have the same temperature to begin with, the molecules inside the box have different velocities. The demon opens a small hole in the barrier only when a faster-than-average molecule arrives from the left part of the box, allowing it to pass to the right part, and when a slower-than-average molecule arrive from the right part of the box, allowing it to pass to the left part. By doing this, the demon causes molecules of higher energy to concentrate in the right part of the box and those of lower energy to concentrate in the left part. This causes the right part to heat up and the left part to cool down, thus enabling work extraction when the system returns to equilibrium in apparent contradiction to the second law of thermodynamics. This experiment shows how information on the speed and location of individual molecules can be transformed into extracted energy, setting the basis for what is now known as ``information engines".

Extensive research and debate has centered around Maxwell's demon since its inception, expanding the concept to more general cases of feedback control based on measurements. It was shown that, for a system with finite memory, the cost of bits erasure nullifies any gain from such a demon \cite{Brillouin1951, Landauer1961, Bennett1987, Mandal2012, Mandal2013, berut2012}. However, it was not until recently that Sagawa and Ueda reached a general upper bound on the amount of work that can be extracted \cite{Sagawa2008, Sagawa2010}, which was also demonstrated experimentally \cite{toyabe2010, Koski2015}. That upper bound was found to be closely related to Shannon's mutual information, which inspired us to look into a possible relation to problems in information theory, a relation that has not yet been explored in full.

Gambling is another field where bits of information were given concrete value, through the analysis of optimal gambling strategies using tools from information theory, an analysis that was first done by Kelly \cite{Kelly1956}. The setting consisted of consecutive bets on some random variable, where all the money won in the previous bet is invested in the current one. Kelly showed that maximizing over the expectation of the gambler's capital would lead to the loss of all capital with high probability after sufficiently many rounds. However, this problem is resolved when maximization is done over the expectation of the logarithm of the capital. Moreover, the logarithm of the capital is additive in consecutive bets, which means that the law of large numbers applies. Under these assumptions, the optimal betting strategy is to place bets proportional to the probability of each result, a strategy referred to as the ``Kelly strategy". Kelly also showed that, given some side information on the event, the profit that can be made compared to the one with no side information is given by Shannon's mutual information. This serves as another hint at a possible relation between information engines and gambling, as the amount of work that can be extracted using measurements, compared to that which can be extracted without measurements, is also given by mutual information.

In this paper, we present an analogy between the analysis of feedback controlled physical systems and the analysis of gambling in information theory\footnote{In \cite{Hirono2015}, an analysis of the gambling problem was carried out using tools from feedback controlled systems and was related to fluctuation theorems in non-equilibrium statistical mechanics (in particular, the Jarzynski equality).}. We show that the optimal control protocol in various systems is analogous to the Kelly strategy, which is the optimal betting strategy. Furthermore, the amount of work extracted after $n$ cycles of an information engine is shown to be analogous to the capital gained after $n$ rounds of gambling. The analogy is then shown on two models: the Szilard Engine, where the particle's location is a discrete random variable, and a particle in some potential field, where the location can be a continuous random variable. The latter prompts us to consider an extension of Kelly gambling to cases with continuous-valued random variables, which is shown to be useful for investment in currency exchange rates or in the stock market using binary options.

This analogy enables us to develop a simple criterion to determine the best control protocol in cases where an optimal protocol is inapplicable and an optimal protocol when the probabilities governing the system are not known. Moreover, well known results for gambling with memory and causal knowledge of side information are applied in the field of physical systems with memory, yielding the optimal control protocol for a certain class of such systems. Under slightly different assumptions, Sagawa and Ueda have derived an upper bound for all such systems in \cite{Sagawa2012}. Throughout this paper, we will ignore the cost of bits erasure, essentially assuming an infinite memory.

The remainder of the paper is organized as follows: in Section \ref{sec:gambling}, we review the problem of horse race gambling, including the optimal betting strategy and maximal gain from side information. In Section \ref{sec:szilard}, we review the operation of the Szilard Engine, its optimal control protocol and maximal work extraction. Then, in Section \ref{sec:analogy}, we present the analogy between these two problems and discuss briefly the implications of such an analogy. In Section \ref{sec:continuous}, we review the mechanism for work extraction from a particle in an external potential field, and present the extension of Kelly gambling to cases with continuous-valued random variables which arises from that physical system. In Section \ref{sec:consequences}, we discuss in more detail some of the implications and uses of the analogy. Finally, in section \ref{sec:otherspec} we discuss speculated analogies to other problems in information theory, and their shortcomings compared to the analogy with gambling.

\section{The Horse Race Gambling}\label{sec:gambling}
The problem of gambling, as presented in \cite{Kelly1956} and \cite{Cover1991}, consists of $n$ experiments whose results are denoted by the random vector $X^n = (X_1,\dots,X_n)$, e.g., the winning horse in $n$ horse races. We are concerned with the case where the gambler has some side information about the races, denoted $Y^n=(Y_1,\dots,Y_n)$. The following notation is used:
\begin{itemize}
  \item $P_X$ - the probability vector of $X$, the winning horse.
  \item $P_{X,Y}$ - the joint probability of $X$ and $Y$.
  \item $P_{X|Y}$ - the conditional probability of $X$ given $Y$.
  \item $P_{X|y}$ - the probability vector of $X$ given an observation $Y=y$ of the side information.
  \item $b_{X|Y}$ - the betting strategy on $X$ given $Y$, describing the fraction of the gambler's capital invested in each horse.
  \item $b_{X|y}$ - a vector describing the betting strategy for $Y=y$.
  \item $o_X$ - a vector describing the amount of money earned for each dollar invested in the winning horse, for each horse.
  \item $S_n$ - the gambler's capital after $n$ experiments.
\end{itemize}
$P_X(x)$, $P_{X,Y}(x,y)$ and $P_{X|Y}(x|y)$ denote the probability mass function (PMF) of $X$, the joint PMF of $X$ and $Y$ and the conditional PMF of $X$ given $Y$, respectively, for the observations $x$ and $y$. Similarly, $b_{X|Y}(x|y)$ and $o_X(x)$ denote the fraction of capital invested and odds, respectively, for $X=x$ and $Y=y$. Unless stated otherwise, we assume $\{(X_i, Y_i)\}_{i=1}^n$ are i.i.d, i.e., $P_{X^n,Y^n}(x^n,y^n) = \prod_{i=1}^n P_{X,Y}(x_i,y_i)$, and that the gambler invests all of his capital in each round.

Without loss of generality, we will set $S_0 = 1$, namely, the gambling starts with $1$ dollar. $S_n$ is then given by:
\begin{equation}\label{DiscreteSn}
S_n = \prod_{i=1}^n b_{X|Y}(X_i|Y_i) o_X(X_i),
\end{equation}
and maximization will be done on $\log S_n$. We define the profit at round $i$ as
\begin{equation}\label{SingleRoundProfit}
\log S_i - \log S_{i-1} = \log\left[b_{X|Y}(X_i|Y_i) o_X(X_i)\right].
\end{equation}
The wealth growth rate is defined as
\begin{equation}
\mathcal{W} = \frac{1}{n} E[\log S_n],
\end{equation}
where the expectation is with respect to $P_{X^n,Y^n}$. The maximal wealth growth rate will be denoted as $\mathcal{W}^*$.

Since the races are assumed i.i.d., the same betting strategy will be used in every round, i.e., $b_{X_i|Y_i} = b_{X|Y}$ for all $i$. As shown in \cite[Chapter~6]{Cover1991}, the optimal betting strategy is given by:
\begin{equation}\label{MaxbX}
b^*_{X|Y} = \arg\max_{b_{X|Y}} E[\log S_n] = P_{X|Y}.
\end{equation}
Substituting $b^*_{X|Y}$ into (\ref{DiscreteSn}) yields the following formula for $\mathcal{W}^*$:
\begin{equation}\label{MaxSnGivenY}
\mathcal{W}^* = \sum_{x,y} P_{X,Y}(x,y) \log\left[P_{X|Y}(x|y)o_X(x)\right].
\end{equation}

As defined in \cite{Kelly1956}, the bet is ``fair" if $o_X(x) = 1/P_X(x)$. It can be seen from (\ref{MaxSnGivenY}) that without side information, no money can be earned in that case. For a fair bet where side information is available, (\ref{MaxSnGivenY}) can be written as
\begin{equation}\label{SnEqI}
\mathcal{W}^* = I(X;Y),
\end{equation}
where $I(X;Y)$ is Shannon's mutual information\footnote{Following the customary notation conventions in information theory, I(X;Y) should not be understood
as a function I of the random outcomes of X and Y, but as a functional of the joint probability distribution of X and Y.} given by:
\begin{equation}
I(X;Y) = \sum_{x,y} P_{X,Y}(x,y) \log\frac{P_{X,Y}(x,y)}{P_X(x) P_Y(y)}.
\end{equation}
In this paper, we are mostly concerned with fair bets. Another point of interest is a constrained bet, which is a fair bet where for each $y$ $b_{X|y}$ is limited to some set $\mathcal{B}$ of possible vectors. For such a bet, the maximal wealth growth rate is given by:
\begin{eqnarray}\label{MaxConstrainedSn}
\mathcal{W}^* &= \max_{b_{X|Y} \in \mathcal{B}} \frac{1}{n} E[\log S_n] \nonumber \\
&= I(X;Y) - \sum_{y \in \mathcal{Y}} P_Y(y) \min_{b_{X|y} \in \mathcal{B}} D(P_{X|y} || b_{X|y}),
\end{eqnarray}
where $D(\cdot||\cdot)$ is the Kullback-Leibler divergence given by:
\begin{equation}
D(P_{X|y} || b_{X|y}) = \sum_x P_{X|y}(x) \log\frac{P_{X|y}(x)}{b_{X|y}(x)},
\end{equation}
and for each $y$ the optimal $b_{X|y} \in \mathcal{B}$ is the one that minimizes $D(P_{X|y} || b_{X|y})$.

\section{The Szilard Engine}\label{sec:szilard}
\begin{figure}[b]
  \centering
  \includegraphics[width=0.5\textwidth]{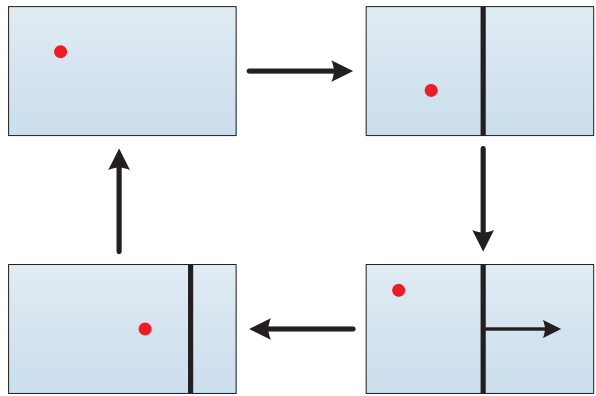}
  \caption{The cycle of the Szilard Engine, starting at the upper left corner.}
  \label{fig:SzilardFigure}
\end{figure}

We now describe the Szilard Engine \cite{Szilard1929}, which involves a single particle of an ideal gas enclosed in a box of volume $V$ and attached to a heat bath of temperature $T$. The engine's cycle consists of the following stages (see Fig. \ref{fig:SzilardFigure}):
\begin{enumerate}
\item The particle moves freely in equilibrium with the heat bath.
\item A divider is inserted, dividing the box into two parts of volumes $V_0^L$ and $V_0^R$ ($V_0^L + V_0^R = V$). The part of the box that contains the particle is denoted by $X$, with the alphabet $\mathcal{X} = \{L, R\}$.
\item A noisy measurement of the particle's location is made; the result is denoted $Y$ with $\mathcal{Y} = \{L, R\}$.
\item Given $Y=y$, the divider is moved quasi-statically until the volumes of the parts are set to the prescribed volumes $V_f^L(y)$ and $V_f^R(y)$.
\item The divider is removed from the box.
\end{enumerate}

Denote the initial normalized volume as $v_0 (x)$, which equals $V_0^L/V$ for $x=L$ and $V_0^R/V$ otherwise. Similarly, the final normalized volume $v_f (x|y)$ is equal to $V_f^L(y)/V$ for $x=L$ and $V_f^R(y)/V$ otherwise. Since the particle starts each cycle in equilibrium with its environment, different cycles of the engine are independent of each other. Moreover, since the particle has no potential energy, the particle's location has a uniform distribution across the volume of the box. Thus, assuming $v_0(x)$ to be the same for each cycle, $X^n$ are i.i.d. with $P_X(x) = v_0(x)$. Following the analysis in \cite{Sagawa2012}, the work extracted for $Y=y$ is given by:
\begin{equation}\label{SingleRoundWork}
W = k_B T \ln \frac{v_f(X|Y)}{P_X(X)},
\end{equation}
where $k_B$ is the Boltzmann constant. It was also shown in \cite{Sagawa2012} that, for every $y \in \mathcal{Y}$, the optimal $v_f$ is
\begin{equation}\label{MaxVf}
v_f^*(\cdot|y) = \arg\max_{v_f} E[W|Y=y] = P_{X|Y}(\cdot|y),
\end{equation}
and the maximal amount of work extracted after $n$ cycles is
\begin{eqnarray}\label{WnEqI}
\max_{v_f} E[W_n] &= n k_B T E\left[\ln\frac{P_{X|Y}(X|Y)}{P_X(X)}\right] \nonumber \\
&= n k_B T I(X;Y).
\end{eqnarray}

Note that the initial location of the barrier $v_0(x)$ can also be optimized, leading to the following formula:
\begin{equation}\label{MaxDiscreteWork}
\max_{v_f, v_0} E[W_n] = n k_B T \max_{P_X} I(X;Y).
\end{equation}

\section{Analogy}\label{sec:analogy}

\begin{table}[bt]
\renewcommand{\arraystretch}{1.0}
\caption{Analogy of gambling and the Szilard Engine}
\label{tab:discrete_equiv}
\centering
\begin{tabularx}{\linewidth}{ | X | X | }
\hline
\textbf{Gambling} & \textbf{The Szilard Engine} \\ \hline
$X_i$ - result of horse race in round $i$. & $X_i$ - location of the particle in cycle $i$. Namely, left or right. \\ \hline
Side information. & Measurement results, possibly with noise. \\ \hline
$Y_i$ - some side information on round $i$. & $Y_i$ - noisy measurement of the particle's location in cycle $i$. \\ \hline
$P_X$ - probability vector of the result. & $P_X$ - probability vector of the particle's location. \\ \hline
$P_{X|y}$ - probability vector of the result given side information $y$. & $P_{X|y}$ - probability vector of the particle's location given measurement $y$. \\ \hline
$o_X(x)$ - amount of money earned for every dollar gambled. & $1/v_0(x)$ -  the reciprocal of the initial volume of the box's parts. \\ \hline
Placing bets on different horses. & Moving the dividers to their final positions. \\ \hline
Choosing the optimal race to bet on. & Choosing the optimal initial location for the divider. \\ \hline
$b_{X|y}(x)$ - amount of money gambled on each result, given $y$. & $v_f (x|y)$ - the normalized final volume of the box's parts, given $y$. \\ \hline
Logarithm of the capital. & Extracted work. \\ \hline
$\log S_n$ - log of the acquired money after $n$ rounds of gambling. & $W_n/(k_B T)$ - total work extracted after $n$ cycles of the engine. \\ \hline
Transforming bits to wealth. & Transforming bits to energy. \\ \hline
(\ref{SingleRoundProfit}), (\ref{MaxbX}), (\ref{SnEqI}) - Profit in round $i$, optimal betting strategy and maximum profit. & (\ref{SingleRoundWork}), (\ref{MaxVf}), (\ref{WnEqI}) - Work extracted in round $i$, optimal control protocol and maximum work extraction. \\ \hline
\end{tabularx}
\end{table}

An analogy between the Szilard Engine and gambling arises from this analysis, as presented in Table \ref{tab:discrete_equiv}. The equations defining both problems, (\ref{SingleRoundProfit}) and (\ref{SingleRoundWork}), are the same if one renames $b_{X|Y}$ as $v_f$ and $o_X$ as $1/P_X$. The analogy also holds for the optimal strategy in both problems, presented in (\ref{MaxbX}) and (\ref{MaxVf}), and the maximum gain, presented in (\ref{SnEqI}) and (\ref{WnEqI}), where $\log S_n$ is renamed $W_n/k_B T$.

Analogous to the way bits of side information are converted into wealth in gambling, bits of measurements are converted into work in the Szilard Engine. Moreover, the actions of the controller in the Szilard Engine are analogous to a gamble on the location of the particle. More specifically, in both problems the goal is to allocate a limited resource (box's volume, gambler's capital) in a way that maximizes the gain (extracted work, increase in capital).

Specifically, the Szilard Engine is analogous to a fair bet, since $v_0(x) = P_X(x)$ and this is analogous to $o_X(x) = 1/P_X(x)$. As stated previously, in a fair bet no money can be earned without side information. In an analogous manner, no work can be extracted from the Szilard Engine without measurements; this conforms with the second law of thermodynamics. However, the option to maximize over $P_X$ in the Szilard Engine has no analogy in gambling as formulated in \cite{Kelly1956} and \cite{Cover1991}. This prompts us to consider an extension to horse race gambling, where the gambler can choose between several different race tracks. This means that (\ref{SnEqI}) can be maximized over all distributions $\{P_X\}$ across some set of distributions $\mathcal{P}$, yielding
\begin{equation}
\mathcal{W}^* = \max_{P_X \in \mathcal{P}} I(X;Y).
\end{equation}

The presented analogy allows us to quantify the loss of energy due to use of a less-than-optimal control protocol in the Szilard engine. Suppose the controller is unable to move the divider to its optimal final position on each round, and is instead limited to some set of divider positions, e.g., the position is limited to certain notches where the divider can be stopped. The maximal extracted work in such a setting is given by the following analogous version of (\ref{MaxConstrainedSn}):
\begin{equation}
\max_{v_f \in \mathcal{V}} E[W_n] = n k_B T I(X;Y) - \sum_{y \in \mathcal{Y}} P_Y(y) \min_{v_f(\cdot|y) \in \mathcal{V}} D(P_{X|y} || v_f(\cdot|y)),
\end{equation}
where $\mathcal{V}$ is the set of allowed partitions of the box. The loss of energy on each cycle due to this limitation is seen to be $D(P_{X|y} || v_f(\cdot|y))$ and the optimal control protocol for each $y$ will be the one that minimizes this loss.

Using the analogy, the Szilard Engine can also be extended to a configuration with multiple dividers. The dividers are inserted simultaneously, dividing the box into $m$ parts with the normalized volume of the $i$th part denoted by $v_0 (i)$. The alphabet of $X$ and $Y$ is then given by $\mathcal{X} = \mathcal{Y} = \{1,...,m\}$. This setting is analogous to a horse race with $m$ horses, where the optimal betting strategy is given by (\ref{MaxbX}). Thus, the optimal control protocol for a measurement $y$ will consist of moving the dividers quasi-statically until for every $i$ the normalized volume of the $i$th part is $v_f^*(i|y) = P_{X|Y}(i|y)$. The work extracted by this scheme, when $v_0(i)$ is optimized, is given by:
\begin{eqnarray}
\max_{v_0(i)} E[W_n] &= n k_B T \max_{P_X} \sum_{i,j=1}^m P_Y(i)P_{X|Y}(j|i)\ln\frac{v_f^*(j|i)}{v_0 (j)} \nonumber \\
&= n k_B T \max_{P_X} \sum_{i,j=1}^m P_Y(i)P_{X|Y}(j|i)\ln\frac{P_{X|Y}(j|i)}{P_X(j)} \nonumber \\
&= n k_B T \max_{P_X} I(X;Y),
\end{eqnarray}
i.e., the amount of work extracted is equal to the maximum given in \cite{Sagawa2008, Sagawa2010}.

\section{A Particle in an External Potential Field and Continuous-Valued Gambling}\label{sec:continuous}
In this section, we review the optimal control protocol for work extraction from a single particle in an external potential field. This case is similar to the Szilard Engine in that for both cases the optimal protocol depends solely on $P_{X|Y}$. Moreover, in both cases the protocol consists of making changes to the system that cause the probability distribution of $X$ after the change to be equal to $P_{X|y}$, where $y$ is the measurement result.

We show that the analogy derived in the previous section holds for this case as well. Since the location of the particle in this case can be a continuous random variable, it prompts us to consider an extension of Kelly gambling to scenarios with continuous random variables. This extension is then shown to describe investment in the stock market using options.

\subsection{A Particle in an External Potential Field}
We now consider an overdamped Langevin system of one particle with the Hamiltonian:
\begin{equation}
H(x,p) = \frac{p^2}{2M} + \mathcal{E}_0(x),
\end{equation}
where $p$ is the particle's momentum, $M$ is its mass, $x$ is its location and $\mathcal{E}_0(x)$ is the potential energy. Again, the particle is kept at constant temperature $T$. The probability distribution of $X$ is then the Boltzmann distribution that arises from $\mathcal{E}_0(x)$ and is denoted $P_X$. Namely,
\begin{equation}
P_X(x) = \frac{1}{Z_0} \exp\left(-\frac{\mathcal{E}_0(x)}{k_B T}\right),
\end{equation}
where $Z_0$ is the partition function, given by:
\begin{equation}
Z_0 = \sum_{x \in \mathcal{X}} \exp\left(-\frac{\mathcal{E}_0(x)}{k_B T}\right).
\end{equation}

For now, we will limit ourselves to cases where $P_{X|y}$ is a Boltzmann distribution for every $y \in \mathcal{Y}$, a constraint that will be relaxed later on. This happens, for example, in the Gaussian case, where $X \sim \mathcal{N}(0,k_B T \sigma_X^2)$ and $Y = X + N$, where $N \sim \mathcal{N}(0, k_B T \sigma_N^2)$ and is independent of $X$, similarly to \cite{Abreu2011}. The optimal control protocol for such a system was found in \cite{HorowitzParrondo2011} and \cite{EspositoBroeck2011} to be as follows:
\begin{itemize}
\item Based on the measurement $y$, $\mathcal{E}_0(x)$ is instantaneously modified by the controller to a different potential field. It then follows that the Boltzmann distribution of $X$ changes from $P_X$ to $Q_{X|y}$, which for every $y$ is the probability distribution of $X$ chosen by the controller. The optimal final distribution, denoted $Q_{X|y}^*$, was shown in \cite{HorowitzParrondo2011} and \cite{EspositoBroeck2011} to be equal to the conditional distribution of $X$ given $y$, i.e., $Q_{X|y}^* = P_{X|y}$.
\item The potential is changed back to $\mathcal{E}_0(x)$ quasi-statically.
\end{itemize}
Noting that $v_f^*$ as presented in (\ref{MaxVf}) is equal to $Q_{X|y}^*$, one notices that both in this case and in the Szilard Engine the optimal control protocol is defined by $P_{X|Y}$. Furthermore, (\ref{SingleRoundWork}) is also valid for this case, with $v_f$ replaced by $Q_{X|y}$. If $X$ is a continuous random variable, $P_X(x)$, $P_{X|Y}(x|y)$ and $Q_{X|y}(x)$ will be the particle's probability density function (PDF), conditional PDF and the PDF chosen by the controller, respectively.

The protocol presented above is optimal in the sense that it attains the upper bound on extracted work, i.e., the extracted work using this protocol with $Q_{X|y} = Q_{X|y}^*$ is given by:
\begin{equation}\label{ContinuousWork}
E[W_n] = n k_B T I(X;Y).
\end{equation}
If the controller controls $\mathcal{E}_0(x)$ as well, the expression in (\ref{ContinuousWork}) can be maximized over all distributions $\{P_X\}$. However, it is important to note that there will always be some constraint over $P_X$, due to the finite volume of the system or due to the method of creating the external potential, or both. Thus, denoting by $\mathcal{P}$ the set of allowed initial distributions $P_X$, the maximal amount of extracted work is given by:
\begin{equation}\label{MaxContinuousWork}
E[W_n] = n k_B T \max_{P_X \in \mathcal{P}} I(X;Y).
\end{equation}
Another point of interest is that setting $Q_{X|y}^* = P_{X|y}$ is not necessarily possible, e.g., if for some values of $y$, $P_{X|y}$ is not of the form
\begin{equation}
P_{X|y}(x) = \frac{1}{Z} \exp\left(-\frac{f(x,y)}{k_B T}\right),
\end{equation}
and thus not a Boltzmann distribution. This gives rise to the following, more general, formula:
\begin{equation}\label{MaxConstrainedWork}
E[W_n] = n k_B T \max_{P_X \in \mathcal{P}} \{ I(X;Y) - \sum_{y \in \mathcal{Y}} P_Y(y) \min_{Q_{X|y} \in \mathcal{P}_B} D(P_{X|y}||Q_{X|y}) \},
\end{equation}
where $\mathcal{P}_B$ is the set of all possible distributions $P_X$ that stems from the set of all possible potentials. Thus, for every $y$, the optimal $Q_{X|y} \in \mathcal{P}_B$ is the one that minimizes $D(P_{X|y}||Q_{X|y})$. Notice that this analysis holds for both continuous and discrete random variables $X,Y$.

It follows that the analogy presented in Table \ref{tab:discrete_equiv} can be extended to work extraction from a particle in an external potential. Again, this system is analogous to a fair bet, in conformance with the second law of thermodynamics. This system is also analogous to a constrained bet, as can be seen from (\ref{MaxConstrainedWork}) and its analogy with (\ref{MaxConstrainedSn}). If $X$ is continuous, an interesting extension to the gambling problem arises where the bet is on continuous random variables. We will now present this extension in detail.

\subsection{Continuous-Valued Gambling}
We consider a bet on some continuous-valued random variable, where the gambler has knowledge of side information. The gambler's wealth is still given by (\ref{DiscreteSn}), but the betting strategy, $b_{X|y}(x)$, and the odds, $o_X(x)$, are functions of the continuous variable $x$ instead of vectors. In the case of stocks or currency exchange rates, for instance, such a betting strategy and odds can be implemented using options\footnote{This is not to be confused with \cite{Erkip1998}, where an investment in $m$ stocks was formulated as a bet on $m$ continuous-valued random variables, and $b$ was a vector of length $m$ denoting the amount invested in each stock.}.

The constraint that the gambler invests all his capital in each round is translated in this case to the constraint
\begin{equation}\label{ContGambConstraint}
\int_\mathcal{X} b_{X|y}(x)dx = 1 \ \ \forall y \in \mathcal{Y}.
\end{equation}
The optimal betting strategy is then given by $b^*_{X|Y}(x|y) = f_{X|Y}(x|y)$, where $f_{X|Y}(x|y)$ is the conditional PDF of $X$ given $Y$, and the bet is said to be fair if $o_X(x) = 1/f_X(x)$, where $f_X(x)$ is the PDF of $X$. For a fair bet, (\ref{SnEqI}) holds and (\ref{MaxConstrainedSn}) holds with the sum replaced by an integral and each probability mass function replaced by the appropriate PDF.

As an example, consider the price of some stock or currency exchange rates, which are continuous-valued. In order to gamble using this model, \emph{binary put options} are used. A binary put option is defined by its \emph{strike price} - if the price of the stock at the expiry date, denoted $X$, is below the strike price, the option pays $1$ dollar and otherwise it is worthless. The investment strategy will consist of selling a binary put with strike price $K$ and buying a binary put with strike price $K + \Delta$ for some $\Delta > 0$; a combination denoted as \emph{a spread} on $(K,K + \Delta]$, which yields $1$ dollar if $x \in (K,K+\Delta]$ and is worthless otherwise. This strategy will be shown to conform with the model of horse race gambling on a continuous-valued random variable.

First, note that the price of a binary option depends on its strike price. We assume it is linear in the interval of a spread, which means that the price of a spread on $(K,K + \Delta]$ is $C_K \Delta$, where $C_K$ is the slope of the option's price in that interval. This assumption is valid for sufficiently small values of $\Delta$. It follows that the gain per dollar for such a spread is
\begin{equation}
o_{K,\Delta} = \frac{1}{C_K \Delta}.
\end{equation}

In order to invest, divide the $x$ axis into $N$ intervals $(K_j, K_j + \Delta_j]$, $j \in \{1, ... , N\}$, where $K_{j+1} = K_j + \Delta_j$ and $N$ can be arbitrarily large. For each interval $j$, denote by $\varphi_j(y)$ the fraction of the capital invested in buying spreads on that interval, given side information $y$. The betting strategy $b_{X|Y}$ is then set as
\begin{equation}
b_{X|Y}(x|y) = \sum_{j=1}^N \frac{\varphi_j(y)}{\Delta_j} \mathbbm{1}_{(K_j,K_j+\Delta_j]}(x),
\end{equation}
where $\mathbbm{1}_{\{\cdot\}}$ is the indicator function. Then, for all $y \in \mathcal{Y}$, the constraint on $b_{X|y}$ is
\begin{equation}
\int\limits_{-\infty}^{\infty} b_{X|y}(x) dx = \sum_{j=1}^N \varphi_j(y) = 1,
\end{equation}
which is the same as the constraint in (\ref{ContGambConstraint}). Similarly, $o_X(x)$ is the piecewise constant function
\begin{equation}\label{OptionOdds}
o_X(x) = \sum_{j=1}^N \Delta_j \cdot o_{K_j,\Delta_j} \cdot \mathbbm{1}_{(K_j,K_j+\Delta_j]}(x).
\end{equation}
The capital at the end of the $i$th round is then given by the capital invested in the spread containing $x_i$ times the gain per dollar for this spread, i.e.,
\begin{eqnarray}
S_i &= \sum_{j=1}^N \varphi_j(y_i) \cdot S_{i-1} \cdot o_{K_j,\Delta_j} \cdot \mathbbm{1}_{(K_j,K_j+\Delta_j]}(X_i) \nonumber \\
&= b_{X|Y}(X_i|y_i) o_X(X_i) S_{i-1}.
\end{eqnarray}
It follows that (\ref{DiscreteSn}) holds. The analogous case to the maximization on $P_X$ in (\ref{MaxContinuousWork}), in this case, is choosing the options to invest in.

For $\Delta \rightarrow 0$, the betting strategy $b_{X|y}(x)$ is not necessarily piecewise constant. Denoting the slope of the option's price as $C(x)$, (\ref{OptionOdds}) is then rewritten as
\begin{equation}
o_X(x) = \frac{1}{C(x)},
\end{equation}
which is also not piecewise constant. It follows that, in this case, the bet is fair if the price of an option with striking price $x$ is $\Pr(X \leq x)$. Also note that, in this limit, the term $I(X;Y)$ in (\ref{SnEqI}) is generally not bounded, as is the case in (\ref{ContinuousWork}) when the particle's location is not discrete. This means that the gain from knowledge of a stock's exact price at a future date is unlimited, similar to the unlimited work extracted from knowledge of the particle's exact location.

We conclude that two often-discussed schemes of work extraction are analogous to the well-known problem of horse race gambling or to the extension of that problem to the continuous-valued case, an extension that actually arose from the analogy. We will now discuss some of the possible benefits from this analogy.

\section{Consequences of the Analogy}\label{sec:consequences}
The analogy that was shown in this paper enables us to use well-known methods and results from horse race gambling to solve problems regarding measurement-based work extraction, and vice versa. Two such cases have already been shown: the Szilard Engine with multiple dividers and continuous-valued gambling. In this section, we present three more problems solved using the analogy: maximum work extraction when the joint distribution of $X$ and $Y$ is unknown, work extraction when some energy is lost in each cycle, e.g., due to friction, and an analysis of systems with memory. In all three cases, the analogy enables us to use known results to gain new insight.

In this section, we assume the control protocol is defined by a probability distribution $Q_{X|y}$, chosen by the controller. Since for every choice of $v_{f}(x|y)$, which defines the control protocol in the Szilard Engine, the following holds
\begin{eqnarray}
v_{f}(x|y) &\in [0,1] \ \ \forall x \in \mathcal{X} \nonumber \\
\sum_{x \in \mathcal{X}} v_f(x|y) &= 1,
\end{eqnarray}
and since (\ref{SingleRoundWork}) holds for both problems, the analysis done henceforth for $Q_{X|y}$ is applicable for $v_{f}(x|y)$ as well.

\subsection{Universal Work Extraction}
In both control protocols presented so far, in order to achieve the upper bound of $E[W] = k_B T I(X;Y)$, it was necessary to know the conditional distribution $P_{X|Y}$ in advance. The question then arises whether this bound could also be achieved when the conditional probability is not known, e.g., a system with an unknown measurement error. The analogous problem in gambling was solved by Cover and Ordentlich \cite{Cover1996} for the case of portfolio management.

Portfolio management is an extension of horse race gambling, where instead of multiple horses with only one winner, the gambler invests in multiple stocks, each performing differently. Namely, following the notation in \cite{Cover1996}, $\mathbf{x_i}$ is a vector representing the price of each stock at time $i$ relative to its price at time $i-1$. $\mathbf{b}$ will denote the portfolio, i.e., a vector whose $j$th element is the fraction of the investors capital invested in the $j$th stock. The investor's capital at time $n$, $S_n$, is then given by $S_{n-1}$ times the vector product of $\mathbf{x_i}$ and $\mathbf{b}$. Alternatively, it can be written as:
\begin{equation}\label{UniversalSn}
S_n = \prod_{i=1}^n \mathbf{b}^t(y_i) \cdot \mathbf{x_i},
\end{equation}
where the notation $\mathbf{b}(y_i)$ represents the fact that the portfolio can depend on side information.

In \cite{Cover1996}, the \emph{$\mu$-weighted universal portfolio with side information} was devised, and was shown to asymptotically achieve the same wealth as the best constant betting strategy for any pair of sequences $x^n, y^n$. Namely, it was shown that
\begin{equation}\label{UniversalBetOptimality}
\lim_{n \rightarrow \infty} \max_{\mathbf{x}^n,y^n} \frac{1}{n} \log\frac{S^*_n(\mathbf{x}^n|y^n)}{\hat{S}_n(\mathbf{x}^n|y^n)} = 0,
\end{equation}
where $\hat{S}_n$ is the wealth achieved by the universal portfolio and $S^*_n$ is the maximal wealth that can be achieved by a constant portfolio, i.e., where $\mathbf{b_i(y_i)} = \mathbf{b(y_i)}^*$ for all $i$. The universal portfolio at time $i$ will be denoted by $\hat{\mathbf{b_i}}(y^i,\mathbf{x}^{i-1})$, which depends on the investor's causal knowledge.

The universal portfolio was given by:
\begin{equation}
\hat{\mathbf{b_i}}(y^i,\mathbf{x}^{i-1}) = \frac{\int\limits_{\mathcal{B}} \mathbf{b} S_{i-1}(\mathbf{b}|y_i) d\mu(\mathbf{b})}{\int\limits_{\mathcal{B}} S_{i-1}(\mathbf{b}|y_i) d\mu(\mathbf{b})},
\end{equation}
where $\mu$ is a measure that can be chosen by the investor under the constraint $\int_\mathcal{B} d\mu = 1$, $\mathcal{B}$ is the set of all possible portfolios $\mathbf{b}$ and $S_{i-1}(\mathbf{b}|y_i)$ is the wealth acquired using portfolio $\mathbf{b}$ along the subsequence $\{j < i : y_j = y_i\}$, i.e.,
\begin{equation}\label{CapitalUsingb}
S_{i-1}(\mathbf{b}|y) = \prod_{j < i : y_j = y} \mathbf{b}^t \cdot \mathbf{x}_j.
\end{equation}
Choosing $\mu$ to be the uniform (Dirichlet$(1,\dots,1)$) distribution, it was also shown that the wealth achieved by the portfolio can be lower bounded by:
\begin{equation}\label{UniversalGainBound}
\log \hat{S}_n(\mathbf{x}^n|y^n) \geq \log S^*_n(\mathbf{x}^n|y^n) - k(m-1) \log (n+1),
\end{equation}
where $m$ is the length of vector $\mathbf{x}$ and $k$ is the cardinality of $\mathcal{Y}$.

We will now consider the case of horse race gambling. Denote $o_j = o_X(j)$, i.e., the odds of the $j$th horse, and, similarly, $b_j$ denotes the $j$th component of $\mathbf{b}$, i.e., the fraction of the capital invested in the $j$th horse by some betting strategy $\mathbf{b}$. Then, (\ref{CapitalUsingb}) can be rewritten as
\begin{equation}
S_{i-1}(\mathbf{b}|y) = \prod_{j=1}^m (b_j o_j)^{n_i(j, y_i)},
\end{equation}
where $n_i(j, y_i)$ is the number of times $X$ was observed to be $j$ and $Y$ was observed to be $y_i$ before the $i$th cycle, i.e., $n_i(j, y) = |\{l:x_l=j, y_l=y, l<i\}|$. When $\mu$ is the uniform distribution, the universal portfolio is then reduced to the following universal betting strategy for the case of horse race gambling:
\begin{eqnarray}\label{UniversalBettingStrategy}
\mathbf{\hat{b}_{i}}(y^i, x^{i-1}) &= \frac{\int\limits_{0}^1 \int\limits_{0}^{1-b_1} \cdots \int\limits_{0}^{1-\sum_{j=1}^{m-2} b_j} \mathbf{b} \prod_{j=1}^m (b_j o_j)^{n_i(j, y_i)} db_1 db_2 \cdots db_{m-1}}{\int\limits_{0}^1 \int\limits_{0}^{1-b_1} \cdots \int\limits_{0}^{1-\sum_{j=1}^{m-2} b_j} \prod_{j=1}^m (b_j o_j)^{n_i(j, y_i)} db_1 db_2 \cdots db_{m-1}} \nonumber \\
&= \frac{\int\limits_{0}^1 \int\limits_{0}^{1-b_1} \cdots \int\limits_{0}^{1-\sum_{j=1}^{m-2} b_j} (b_1, \dots, b_m) \prod_{j=1}^m b_j^{n_i(j, y_i)} db_1 db_2 \cdots db_{m-1}}{\int\limits_{0}^1 \int\limits_{0}^{1-b_1} \cdots \int\limits_{0}^{1-\sum_{j=1}^{m-2} b_j} \prod_{j=1}^m b_j^{n_i(j, y_i)} db_1 db_2 \cdots db_{m-1}} \nonumber \\
&= \left(\frac{n_i(1, y_i) + 1}{n_i(y_i) + m}, \dots , \frac{n_i(m, y_i)+1}{n_i(y_i) + m}\right),
\end{eqnarray}
where $n_i(y) = |\{l:y_l=y, l<i\}|$ is the number of times $Y$ was observed to be $y$ before the $i$th cycle and $b_m = 1-\sum_{j=1}^{m-1} b_j$.

Using the analogy presented above, this universal portfolio can be adapted straightforwardly into a universal control protocol in cases where $X$ has a finite alphabet. In this control protocol, $Q_{X_i|y^i,x^{i-1}}$ is given by the right-hand-side (RHS) of (\ref{UniversalBettingStrategy}) and the extracted work is lower bounded by:
\begin{equation}
\hat{W}_n \geq W^*_n - k_B T k(m-1)\ln(n+1),
\end{equation}
a bound that follows directly from (\ref{UniversalGainBound}), where $m$ is the cardinality of $\mathcal{X}$ and $k$ is the same as before. Namely, the work extracted by this universal control protocol is asymptotically equal to the work extracted by the best constant control protocol, i.e., the control protocol in which $Q_{X_i|y_i} = Q_{X|y_i}^*$ for all $i$. However, this derivation is applicable only for finite alphabets.

\subsection{Imperfect Work Extraction}
Another outcome that arises from the analogy shown above is the analysis of an imperfect system of work extraction. Consider a system where some amount of energy $f(x)$ is lost in each cycle, e.g., due to friction. The work extracted in each cycle is then given by:
\begin{equation}\label{ImperfectWork}
W = k_B T \ln \frac{Q_{X|Y}(X|Y)}{P_X(X)} - f(X).
\end{equation}
This is analogous to an unfair bet with the odds
\begin{equation}\label{UnfairOdds}
o_X(x) = \frac{1}{P_X(x)} \exp(-f_T(x)),
\end{equation}
where $f_T(x) = f(x)/k_B T$ and $T$ is an ``unfairness" parameter.

As shown in \cite[Chapter~6]{Cover1991}, if the gambler has to invest all the capital in each round, the optimal betting strategy is independent of $o_X(x)$, i.e., for the odds given in (\ref{UnfairOdds}) the optimal betting strategy is still given by (\ref{MaxbX}), which yields
\begin{equation}
E[\ln S_i - \ln S_{i-1}|Y_i=y_i] = D(P_{X|y_i} || P_X) - E[f_T(X_i)|Y_i=y_i].
\end{equation}
However, it may be the case that for some values of $y$ the gambler should not gamble at all. Specifically, in rounds where $Y_i=y_i$ and $D(P_{X_i|y_i}||P_X) \leq E[f_T(X_i)|Y_i=y_i]$, gambling should be avoided.

In the same manner, the optimal control protocol for imperfect systems of work extraction is still given by:
\begin{equation}\label{OptPy}
Q_{X|y}^* = P_{X|y},
\end{equation}
but for some measurement results it may be preferable not to perform the cycle at all. Substituting (\ref{OptPy}) into (\ref{ImperfectWork}) and taking the average w.r.t. $P_{X|y}$ yields
\begin{equation}
E[W|Y=y] = k_B T D(P_{X|y} || P_X) - E[f(X)|Y=y].
\end{equation}
Thus, the engine's cycle should be performed only if $y_i$ satisfies $k_B T D(P_{X_i|y_i}||P_{X_i}) > E[f(X_i)|Y_i=y_i]$.

\subsection{Systems With Memory}\label{sec:memory}
Finally, we would like to analyze cases where the different cycles of the engine, or different measurements, are not independent. An upper-bound for this case was derived in \cite{Sagawa2012}, under different assumptions. Namely, we assume the controller has causal knowledge of previous states of the system, and that the measurement result on each cycle can explicitly depend on previous measurements. Under these assumptions, we derive the optimal control protocol and the general gain in work extraction due to measurements, and in one example also the maximal amount of work that can be extracted.

Again, we use known results from the analysis of gambling on dependent horse races. The gain in wealth due to casual knowledge of side information, as shown in \cite{Permuter2011}, is
\begin{equation}
E[\log S_n(X^n||Y^n)] - E[\log S_n(X^n)] = I(Y^n \rightarrow X^n).
\end{equation}
The term $I(Y^n \rightarrow X^n) \triangleq \sum_{i=1}^n I(X_i;Y^i|X^{i-1})$ is the directed information from $Y^n$ to $X^n$, as defined by Massey \cite{Massey1990}, and $S_n(X^n||Y^n)$ indicates the betting strategy at round $i$ depends causally on previous results $X^{i-1}$ and side-information $Y^i$. The optimal betting strategy in this case is given by $b_{X^n||Y^n}^*(x^n||y^n) = P_{X^n||Y^n}(x^n||y^n)$, where $P_{X^n||Y^n}(x^n||y^n) = \prod_{i=1}^n P_{X_i|Y^i,X^{i-1}}(x_i|y^i,x^{i-1})$ is the causal conditioning of $X^n$ by $Y^n$, as defined by Kramer \cite{Kramer1998, Kramer2003}.

Analogously, in a physical system of work extraction where different cycles are dependent, e.g., when the system does not reach equilibrium between cycles, the formulas presented so far are no longer valid. Instead, the controller's causal knowledge of previous states and measurements can be used, meaning the optimal control protocol is given by
\begin{eqnarray}\label{MaxMemoryVf}
\fl \arg\max_{Q_{X_i|y^i,x^{i-1}}} & E[W_i|Y^i=y^i, X^{i-1}=x^{i-1}] \nonumber \\
&= \arg\max_{Q_{X_i|y^i,x^{i-1}}} k_B T \sum_{i=1}^n E\left[\ln\frac{Q_{X_i|y^i,x^{i-1}}(X_i)}{P_X(X_i)} \bigg| Y^i=y^i,X^{i-1}=x^{i-1} \right] \nonumber \\
&= P_{X_i|y^i,x^{i-1}},
\end{eqnarray}
where for each $(y^i, x^{i-1})$, $Q_{X_i|y^i,x^{i-1}}$ is some probability distribution of $X$ chosen by the controller. This means that maximal work extraction is given by:
\begin{eqnarray}\label{MaxWorkMemory}
E[W_n(X^n||Y^n)] &= k_B T \sum_{i=1}^n E\left[\ln\frac{P_{X_i|Y^i,X^{i-1}}(X_i|Y^i,X^{i-1})}{o_i(X^i)} \right] \nonumber \\
&= -k_B T \sum_{i=1}^n \left[H(X_i|Y^i,X^{i-1}) - O_i\right],
\end{eqnarray}
where $H(\cdot|\cdot)$ is Shannon's conditional entropy given by:
\begin{equation}
H(X|Y) = - \sum_{x,y} P_{X,Y}(x,y) \log P_{X|Y}(x|y),
\end{equation}
$o_i(X^i)$ is some function of the current and previous states, $O_i = E[\ln o_i(X^i)]$ and the notation $W_n(X^n||Y^n)$ indicates that the control protocol at round $i$ depends causally on previous states $X^{i-1}$ and side-information $Y^i$.  Without access to measurement results, which is equivalent to setting $Y_i = \emptyset$ for all $i$, the maximal work extraction is
\begin{eqnarray}\label{MaxWorkMemoryNoY}
E[W_n(X^n)] &= k_B T \sum_{i=1}^n E\left[\ln\frac{P_{X_i|X^{i-1}}(X_i|X^{i-1})}{o_i(X^i)}\right] \nonumber \\
&= -k_B T \sum_{i=1}^n \left[H(X_i|X^{i-1}) - O_i\right].
\end{eqnarray}
Subtracting (\ref{MaxWorkMemoryNoY}) from (\ref{MaxWorkMemory}), the gain in work extraction due to causal knowledge of measurement results is
\begin{equation}\label{WorkGainMem}
E[W_n(X^n||Y^n)] - E[W_n(X^n)] = k_B T I(Y^n \rightarrow X^n),
\end{equation}
analogously to the horse race gamble, where the gain in wealth growth rate due to causal knowledge of side information is $I(Y^n \rightarrow X^n)$.

\begin{example}
\normalfont
Consider the Szilard Engine where the initial placement of the barrier in each cycle is done before the system reaches equilibrium. As a result, in each cycle the particle has a higher probability to be in the same part of the box that it was in in the previous one. Denote by $p$ the probability that the particle moved from one part of the box to the other between cycles and $q$ the probability of measurement error, i.e., $\Pr(X_i \neq X_{i-1}) = p$ and $\Pr(X_i \neq Y_i) = q$ for each $i$. Since the measurement device has no memory and only the previous location affects the current one, the system has the Markov properties $X_i - (X_{i-1}, Y_i) - (Y^{i-1},X^{i-2})$, $X_i - X_{i-1} - X^{i-2}$ and $Y_i - X_i - (X^{i-1},Y^{i-1})$. Equation (\ref{WorkGainMem}) can then be written explicitly as
\begin{eqnarray}
E[W_n(X^n||Y^n)] - E[W_n(X^n)] &= k_B T \sum_{i=1}^n I(X_i;Y^i|X^{i-1}) \nonumber \\
&= k_B T \sum_{i=1}^n \left[ H(X_i|X_{i-1}) - H(X_i|X_{i-1}, Y_{i}) \right] \nonumber \\
&= k_B T \sum_{i=1}^n I(X_i;Y_i|X_{i-1}) \nonumber \\
&= k_B T \sum_{i=1}^n \left[ H(Y_i|X_{i-1}) - H(Y_i | X_i) \right] \nonumber \\
&= k_B T n \left[ H_b(p*q) - H_b(q) \right],
\end{eqnarray}
where $p*q = pq + (1-p)(1-q)$ and $H_b(x) = -x \ln x -(1-x)\ln(1-x)$. For this example, the analogous case is a horse race with two horses, where the probability that a horse would win two consecutive races is $1-p$, and the gambler has side information on the outcome of the race with error probability $q$. In that case, the gain in wealth growth rate due to the side information is given by
\begin{eqnarray}
E[\log(S(X^n||Y^n)] - E[\log S(X^n)] &= I(Y^n \rightarrow X^n) \nonumber \\
&= \sum_{i=1}^n I(X_i;Y^i|X^{i-1}) \nonumber \\
&= n \left[ H_b(p*q) - H_b(q) \right].
\end{eqnarray}
Specifically, the particle remaining in the same part of the box is analogous to a horse winning two consecutive races, and the error in side information is analogous to a measurement error.
\end{example}

\begin{example}\label{ex:Hysteresis} \normalfont
Consider a system of work extraction where the position of the particle in each cycle is independent of previous cycles, but each measurement depends on previous ones, e.g., due to hysteresis. However, both for the Szilard Engine and for a particle in an external potential, the probability distribution at the $i$th cycle can be chosen by the controller. Thus, the controller can introduce a dependence of $X_i$ on $X^{i-1}, Y^{i-1}$ through the choice of $P_{X_i}$. Maximal work extraction is then given by
\begin{eqnarray}\label{MaxWnHisteresis}
E[W_n] &= k_B T \max_{\{P_{X_i|X^{i-1},Y^{i-1}}\}_{i=1}^n \in \mathcal{P}} \sum_{i=1}^n E\left[\ln\frac{P_{X_i|Y^i,X^{i-1}}(X_i|Y^i, X^{i-1})}{P_{X_i|Y^{i-1},X^{i-1}}(X_i|Y^{i-1}, X^{i-1})}\right] \nonumber \\
&= k_B T \max_{\{P_{X_i|X^{i-1},Y^{i-1}}\}_{i=1}^n \in \mathcal{P}} \sum_{i=1}^n I(X_i;Y_i|X^{i-1},Y^{i-1}),
\end{eqnarray}
where $\mathcal{P}$ is the set of possible distributions and $\{P_{Y_i|Y^{i-1},X^i}\}_{i=1}^n$ is a constant of the measuring device.

This example is analogous to gambling where the gambler can choose in each round between several different race tracks, with different tracks independent of each other. In that case, in each round the probability distribution of the horses depends on the chosen track, which itself depends on the causal knowledge the gambler has. Thus, $X_i$ depends on $X^{i-1}, Y^{i-1}$ through the gambler's choice of $P_{X_i}$, i.e., if in round $i$ track $j$ was chosen, and the bet is fair, the odds at round $i$ are
\begin{equation}
o_i(x_i) = \frac{1}{P_j(x_i)} = \frac{1}{P_{X_i|X^{i-1},Y^{i-1}}(x_i|x^{i-1},y^{i-1})},
\end{equation}
where $P_j(x_i)$ is the PMF corresponding to the $j$th track. The maximal wealth growth rate is then given by
\begin{eqnarray}
\mathcal{W}^* &= \max_{\{P_{X_i|X^{i-1},Y^{i-1}}\}_{i=1}^n \in \mathcal{P}} \frac{1}{n} \sum_{i=1}^n E\left[\ln\frac{P_{X_i|Y^i,X^{i-1}}(X_i|Y^i, X^{i-1})}{P_{X_i|Y^{i-1},X^{i-1}}(X_i|Y^{i-1}, X^{i-1})}\right] \nonumber \\
&= \max_{\{P_{X_i|X^{i-1},Y^{i-1}}\}_{i=1}^n \in \mathcal{P}} \frac{1}{n} \sum_{i=1}^n I(X_i;Y_i|X^{i-1},Y^{i-1}),
\end{eqnarray}
where $\mathcal{P}$ is the set of distributions of the different tracks. This equation is analogous to (\ref{MaxWnHisteresis}), and in both cases no gain is possible without knowledge of $Y_i$.

A simpler form of this example would be to assume the same probability distribution of $X$ is chosen for every $i$, so that $X_i$ is independent of $X^{i-1}, Y^{i-1}$. For instance, consider a case where the measurement device can either be in a "good" state, yielding error-free measurements, or a "bad" state, yielding measurements which are independent of $X$, and in each cycle it is likely to remain in the same state as it was in the previous cycle. Thus, even though $X_i$ is independent of $X^{i-1}$ and $Y^{i-1}$, previous states and measurements contain information on the state of the measurement device given $Y_i$. In such a case, $X_i$ are i.i.d. and (\ref{MaxWnHisteresis}) is reduced to
\begin{eqnarray}\label{WnHisteresis}
E[W_n] &= k_B T \sum_{i=1}^n E\left[\ln\frac{P_{X_i|Y^i,X^{i-1}}(X_i|Y^i, X^{i-1})}{P_{X}(X_i)}\right] \nonumber \\
&= k_B T I(Y^n \rightarrow X^n).
\end{eqnarray}
\end{example}

In Example \ref{ex:Hysteresis}, the amount of work that can be extracted is given by (\ref{MaxWnHisteresis}), which is a solution to a maximization problem. Hence, it would be beneficial if the maximized expression was concave.
\begin{lemma}\label{ConcavityLemma}
Let $f(P_{X^n||Y^{n-1}}, P_{Y^n||X^n}) = \sum_{i=1}^n I(X_i;Y_i|X^{i-1},Y^{i-1})$. Then $f$ is concave in $P_{X^n||Y^{n-1}}$ with $P_{Y^n||X^n}$ constant.
\end{lemma}
\textit{Proof:} See the appendix.

Notice that a one-to-one mapping exists between $P_{X^n||Y^{n-1}}$ and $\{P_{X_i|X^{i-1},Y^{i-1}}\}_{i=1}^n$ \cite[Lemma~3]{Permuter2009}. It then follows from Lemma \ref{ConcavityLemma} that the maximization problem in (\ref{MaxWnHisteresis}) can be solved using the tools of convex optimization, if $\mathcal{P}$ is convex. Alternatively, the alternating maximization procedure can be used to maximize over each term $P_{X_i|X^{i-1},Y^{i-1}}$ separately while setting all other terms to be constant, beginning with $i=n$ and moving backward to $i=1$, similarly to \cite{Naiss2013}. Since each term depends only on previous terms and not on the following ones, this procedure will yield the global maximum as needed.

\section{Other Speculated Analogies}\label{sec:otherspec}
Since mutual information appears in another context in information theory, called channel coding, the reader might speculate that an analogy between the Szilard engine and channel coding may be more natural. However, we feel the analogy with gambling is the most fitting and complete.

In channel coding, an encoder needs to encode a message of $nR$ bits into a stream of $n$ bits, denoted $x^n$, which would be transmitted through a channel with output $y^n$ and a probability distribution $P_{Y|X}$. A decoder then needs to reconstruct the original message out of $y^n$. The maximal value of $R$ in this scenario, which could be seen as the maximal gain of message bits, is known as the \emph{channel capacity} and is equal to $I(X;Y)$ \cite[Chapter~7]{Cover1991}. While the maximal gain in channel coding is defined by mutual information, and while $p(x|y)$ plays an important part in the decoding strategy, we feel an analogy of this with the Szilard engine is lacking.

First and foremost, in channel coding, as the decoder attempts to estimate $x^n$ based on $y^n$ it uses the fact that only certain values of $x^n$ are possible. This is not the case in the Szilard engine, nor is it the case in horse race gambling, where estimation of $X$ is performed in each round. As a result, if $X^n$ are i.i.d knowledge of previous values of $X$ does not help the gambler, nor does it help the controller in the Szilard engine, i.e., no gain or work extraction is possible if $Y_i=X_{i-1}$. In channel coding, on the other hand, the capacity of a channel with $Y_i = X_{i-1}$ is the same as that of a perfect channel, where $Y_i = X_i$.

Second, once systems with memory are considered in Section \ref{sec:memory}, the extracted work is characterized by the directed information from $Y^n$ to $X^n$ \cite{Massey1990}, $I(Y^n \rightarrow X^n) = \sum_{i=1}^n I(X_i;Y^i|X^{i-1})$. This is also what characterizes the maximal wealth growth rate in horse race gambling with memory. However, in the setting of channel coding over channels with memory, the capacity is characterized by $I(X^n \rightarrow Y^n) = \sum_{i=1}^n I(Y_i;X^i|Y^{i-1})$. Again, using the previous example where $Y_i=X_{i-1}$, one can see that indeed $I(Y^n \rightarrow X^n) = 0$, but $I(X^n \rightarrow Y^n) > 0$ even though no work extraction is possible.

In conclusion, while there are connections between channel coding and Maxwell's demon \cite{Kafri2012}, we feel that an analogy between the two is lacking in several key aspects, and offers no further insight, compared to the analogy proposed in this paper.

\section{Conclusions}
In this paper we have shown an analogy between the field of gambling in information theory and the analysis of information engines in statistical mechanics. This analogy consisted of a one-to-one mapping of concepts and equations between those two fields, which enabled us to use methods and results from one field to gain new insights in the other. Such insights included universal work extraction, continuous-valued gambling and information engines with memory, among others.

While in this paper we reviewed only two information engines, the analogy is valid for every engine where the optimal control protocol dictates a change of the Boltzmann distribution so that it is equal to $P_{X|y}$ for every measurement $y$. Analysis of other systems could yield further insight into this analogy and, through it, into gambling.

\ack
The authors would like to thank Oleg Krichevsky for valuable discussions.

The work of D. Vinkler and H. Permuter was supported by the Israel Science Foundation (grant no. 684/11) and the ERC starting grant. The work of N. Merhav was supported by the Israel Science Foundation (ISF), grant no. 412/12.

\appendix
\section*{Appendix} \label{sec:ConcavityProof}
\setcounter{section}{1}
In this appendix we prove the concavity of $\sum_{i=1}^n I(X_i;Y_i|X^{i-1},Y^{i-1})$ in $P_{X^n||Y^{n-1}}$ with $P_{Y^n||X^n}$ constant, where $P_{X^n||Y^{n-1}}$ is the causal conditioning given by:
\begin{equation}
P_{X^n||Y^{n-1}}(x^n||y^{n-1}) = \prod_{i=1}^n P_{X_i|X^{i-1},Y^{i-1}}(x_i|x^{i-1},y^{i-1}).
\end{equation}
Namely, we would like to show that for any $\lambda \in [0,1]$ and causal conditioning measures $P^1_{X^n||Y^{n-1}}$ and $P^2_{X^n||Y^{n-1}}$,
\begin{eqnarray}\label{ConcavityDef}
f(\lambda P^1_{X^n||Y^{n-1}} + \bar{\lambda} P^2_{X^n||Y^{n-1}}, P_{Y^n||X^n}) &\geq \lambda f(P^1_{X^n||Y^{n-1}},P_{Y^n||X^n}) \nonumber \\
&+ \bar{\lambda} f(P^2_{X^n||Y^{n-1}},P_{Y^n||X^n}),
\end{eqnarray}
where $\bar{\lambda} = 1-\lambda$ and
\begin{equation}
f(P^j_{X^n||Y^{n-1}},P_{Y^n||X^n}) = \sum_{i=1}^n I_j(X_i;Y_i|X^{i-1},Y^{i-1}),
\end{equation}
where $I_j(X_i;Y_i|X^{i-1},Y^{i-1})$ is the mutual information induced by $P^j_{X^n||Y^{n-1}}$ for $j \in \{1,2\}$.

Let $S \sim \mathcal{B}(\lambda)$. Denote
\begin{eqnarray}
P^1_{X^n||Y^{n-1}}(x^n||y^{n-1}) &= \prod_{i=1}^n P_{X_i|X^{i-1},Y^{i-1},S}(x_i|x^{i-1},y^{i-1},0) \nonumber \\
P^2_{X^n||Y^{n-1}}(x^n||y^{n-1}) &= \prod_{i=1}^n P_{X_i|X^{i-1},Y^{i-1},S}(x_i|x^{i-1},y^{i-1},1).
\end{eqnarray}
It follows that for all $i$
\begin{eqnarray}
I_1 (X_i;Y_i|X^{i-1},Y^{i-1}) &= I(X_i;Y_i|X^{i-1},Y^{i-1},S=0) \nonumber \\
I_2 (X_i;Y_i|X^{i-1},Y^{i-1}) &= I(X_i;Y_i|X^{i-1},Y^{i-1},S=1).
\end{eqnarray}
The RHS of (\ref{ConcavityDef}) emerges from the following derivation:
\begin{eqnarray}\label{ConcavityRhs}
\sum_{i=1}^n I(S, X_i;Y_i|X^{i-1},Y^{i-1}) &\geq& \sum_{i=1}^n I(X_i;Y_i|X^{i-1},Y^{i-1},S) \nonumber \\
&=& \sum_{i=1}^n P(S = 0) I(X_i;Y_i|X^{i-1},Y^{i-1},S = 0) \nonumber \\
&&+ \sum_{i=1}^n P(S = 1) I(X_i;Y_i|X^{i-1},Y^{i-1},S = 1) \nonumber \\
&=& \lambda \sum_{i=1}^n I_1 (X_i;Y_i|X^{i-1},Y^{i-1}) \nonumber \\
&&+ \bar{\lambda} \sum_{i=1}^n I_2 (X_i;Y_i|X^{i-1},Y^{i-1}) \nonumber \\
&=& \lambda f(P^1_{X^n||Y^{n-1}},P_{Y^n||X^n}) \nonumber \\
&&+ \bar{\lambda} f(P^2_{X^n||Y^{n-1}},P_{Y^n||X^n}).
\end{eqnarray}
As for the LHS, notice that
\begin{eqnarray}\label{ConcavityLhsIncomplete}
\fl \sum_{i=1}^n I(S, X_i;Y_i|X^{i-1},Y^{i-1}) &= \sum_{i=1}^n I(X_i;Y_i|X^{i-1},Y^{i-1}) + \sum_{i=1}^n I(S;Y_i|X^i,Y^{i-1}) \nonumber \\
&\stackrel{(a)}{=} \sum_{i=1}^n I(X_i;Y_i|X^{i-1},Y^{i-1}),
\end{eqnarray}
where $(a)$ follows from the fact that $P_{Y^n||X^n}$ is constant and thus the Markov property $Y_i - (X^i,Y^{i-1}) - S$ holds for all $i$. From (\ref{ConcavityRhs}) and (\ref{ConcavityLhsIncomplete}), it follows that for any $\lambda \in [0,1]$
\begin{eqnarray}
\sum_{i=1}^n I(X_i;Y_i|X^{i-1},Y^{i-1}) \geq &\lambda f(P^1_{X^n||Y^{n-1}}, P_{Y^n||X^n}) \nonumber \\
&+ \bar{\lambda} f(P^2_{X^n||Y^{n-1}}, P_{Y^n||X^n}).
\end{eqnarray}
In order to complete the proof of (\ref{ConcavityDef}), it is necessary to show that the RHS of (\ref{ConcavityLhsIncomplete}) is the LHS of (\ref{ConcavityDef}), i.e., it is needed to show that
\begin{equation}\label{ConcavityLhsCompletion}
P_{X^n||Y^{n-1}} = \lambda P^1_{X^n||Y^{n-1}} + \bar{\lambda} P^2_{X^n||Y^{n-1}}.
\end{equation}

\begin{lemma}
For every pair of r.v. vectors $\{X^n, Y^n\}$ and r.v. $S$ that satisfy the Markov property $Y_i - (X^i,Y^{i-1}) - S$,
\begin{equation}
P_{X^n||Y^{n-1}}(x^n||y^{n-1}) = \sum_s P_{S,X^n||Y^{n-1}}(s,x^n||y^{n-1}),
\end{equation}
where $P_{S,X^n||Y^{n-1}} = P_{Z^{n+1}||Y^{n-1}}$ for $Z^{n+1} = \{S,X^n\}$.
\end{lemma}
\textit{Proof:}
\begin{eqnarray}
P_{X^n||Y^{n-1}}(x^n||y^{n-1}) &= \frac{P_{X^n,Y^n}(x^n,y^n)}{P_{Y^n||X^n}(y^n||x^n)} \nonumber \\
&= \frac{\sum_s P_{S,X^n,Y^n}(s,x^n,y^n)}{P_{Y^n||X^n}(y^n||x^n)} \nonumber \\
&\stackrel{(a)}{=} \frac{\sum_s P_{S,X^n||Y^{n-1}}(s,x^n||y^{n-1})P_{Y^n||X^n}(y^n||x^n)}{P_{Y^n||X^n}(y^n||x^n)},
\end{eqnarray}
where $(a)$ follows from the definition of $P_{S,X^n||Y^{n-1}}$, the probability chain rule and the Markov property.

\begin{lemma}
For every pair of r.v. vectors $\{X^n, Y^n\}$ and r.v. $S$,
\begin{equation}
P_{S,X^n,Y^{n-1}}(s,x^n||y^{n-1}) = P_S(s)P_{X^n||S,Y^{n-1}}(x^n||s,y^{n-1}),
\end{equation}
where $P_{X^n||S,Y^{n-1}} = P_{X^n||Z^n}$ for $Z^n = \{S,Y^{n-1}\}$.
\end{lemma}
\textit{Proof:} The proof follows directly from the definitions of $P_{S,X^n||Y^{n-1}}$ and $P_{X^n||S,Y^{n-1}}$.

Equation (\ref{ConcavityLhsCompletion}) emerges from the previous lemmas as follows:
\begin{eqnarray}
P_{X^n||Y^{n-1}}(x^n||y^{n-1}) &=& \sum_s P_S(s)P_{X^n||S,Y^{n-1}}(x^n||s,y^{n-1}) \nonumber \\
&=& \lambda \prod_{i=1}^n P_{X_i|X^{i-1},Y^{i-1},S}(x_i|x^{i-1},y^{i-1},0) \nonumber \\
&&+ \bar{\lambda} \prod_{i=1}^n P_{X_i|X^{i-1},Y^{i-1},S}(x_i|x^{i-1},y^{i-1},1) \nonumber \\
&=& \lambda P^1_{X^n||Y^{n-1}}(x^n||y^{n-1}) + \bar{\lambda} P^2_{X^n||Y^{n-1}}(x^n||y^{n-1}).
\end{eqnarray}

\section*{References}
\bibliographystyle{iopart-num}
\bibliography{IEEEabrv,equivalence}

\end{document}